\documentclass{ws-procs9x6-cpt25}
\begin{document}

\def\thpr{these proceedings}

\def\al{\alpha}
\def\be{\beta}
\def\ga{\gamma}
\def\de{\delta}
\def\ep{\epsilon}
\def\ve{\varepsilon}
\def\ze{\zeta}
\def\et{\eta}
\def\th{\theta}
\def\vt{\vartheta}
\def\io{\iota}
\def\ka{\kappa}
\def\la{\lambda}
\def\vpi{\varpi}
\def\rh{\rho}
\def\vr{\varrho}
\def\si{\sigma}
\def\vs{\varsigma}
\def\ta{\tau}
\def\up{\upsilon}
\def\ph{\phi}
\def\vp{\varphi}
\def\ch{\chi}
\def\ps{\psi}
\def\om{\omega}
\def\Ga{\Gamma}
\def\De{\Delta}
\def\Th{\Theta}
\def\La{\Lambda}
\def\Si{\Sigma}
\def\Up{\Upsilon}
\def\Ph{\Phi}
\def\Ps{\Psi}
\def\Om{\Omega}
\def\mn{{\mu\nu}}

\def\cL{{\cal L}}
\def\lrvec#1{ \stackrel{\leftrightarrow}{#1} }

\def\fr#1#2{{{#1} \over {#2}}}
\def\half{{\textstyle{1\over 2}}}
\def\quar{{\textstyle{1\over 4}}}
\def\eigh{{\textstyle{1\over 8}}}
\def\frac#1#2{{\textstyle{{#1}\over {#2}}}}

\def\prt{\partial}

\def\etal{{\it et al.}}

\def\pt#1{\phantom{#1}}
\def\ol#1{\overline{#1}}

\def\sb{\overline{s}{}}

\def\stx{\sb^{\bar t \bar x}}
\def\sty{\sb^{\bar t \bar y}}
\def\stz{\sb^{\bar t \bar z}}

\newcommand{\beq}{\begin{equation}}
\newcommand{\eeq}{\end{equation}}
\newcommand{\bea}{\begin{eqnarray}}
\newcommand{\eea}{\end{eqnarray}}
\newcommand{\bit}{\begin{itemize}}
\newcommand{\eit}{\end{itemize}}
\newcommand{\rf}[1]{(\ref{#1})}

\def\bt{{\tilde b}}
\def\ct{{\tilde c}}
\def\dt{{\tilde d}}
\def\gt{{\tilde g}}
\def\Ht{{\tilde H}}

\newcommand{\refeq}[1]{(\ref{#1})}
\def\etal {{\it et al.}}

\title{Expanding the Reach of Laboratory SME Searches Using Higher-Precision Boost Transformations}

\author{Jay D. Tasson}

\address{Physics and Astronomy Department, Carleton College,\\
Northfield, Minnesota 55057, United States}

\begin{abstract}
Additional sensitivities to Lorentz violation can be obtained from existing experiments
by considering additional boost-suppressed effects.  The additional Lorentz-violating signals arise as variations in experimental observables at the commonly-used sidereal frequency as well as more novel frequencies.  In this work we provide some examples
that serve to illustrate how interesting signals arise from the structure
of the relevant boost transformations.

\end{abstract}

\bodymatter

\section{Introduction}
The Standard-Model Extension (SME) is a test framework for organizing a systematic search for Lorentz and CPT violation,
perhaps arising from Planck-scale physics.\cite{ks1998,kgrav}
Searches in the context of the SME have now been performed across nearly all areas of physics\cite{datatables} ranging from atomic physics,\cite{berlin}
to gravitational waves,\cite{gws} to collider physics.\cite{collider}
Due to the ubiquity of ordinary matter,
searches with fermions such as protons, neutrons, and electrons
have been a particularly active area.
The complete minimal fermion sector of the SME (operators of mass dimension $d=3$ and $d=4$)
was written down in 1998 by Colladay and Kosteleck\'y.\cite{ks1998}
Since then, dozens of experiments have sought Lorentz and CPT violating effects using fermions.\cite{datatables}
Despite more than 25 years of intense activity,
even the coefficients associated with protons, neutrons, and electrons had not been completely explored as of the publishing of the recent edition of the data tables.\cite{datatables}
Our recent work illustrates that these unconstrained 
coefficients can be explored using higher-precision frame transformations
in ways that both generate new constraints from existing published results and lead to new discovery potential for future analyses.\cite{jof}
While assistance from computer algebra is needed to address the full space of 44 degrees of freedom per fermion that exists in the applicable limit of the minimal SME,
it is instructive to examine simple examples of the general procedure that can be presented 
with a few lines of algebra.
We present such examples in the remainder of this work that illustrate key features of our general results.

\section{The fermion sector}

We begin by presenting the parts of the Lagrange density for the minimal fermion sector that are relevant for our study as a means of defining 
the key variables for the analysis to follow\cite{ks1998}
\beq
\cL = \frac{1}{2} i \overline{\ps} 
(\ga_\nu + d_\mn \ga_5 \ga^\mu 
   + \half g_{\la \mu \nu} \si^{\la \mu})
\lrvec{\prt^\nu} \ps
- \ol{\ps} 
(m + b_\mu \ga_5 \ga^\mu 
   + \half H_\mn \si^\mn)
   \ps.
\label{lagr}
\eeq
The objects 
$b_\mu$, $d_\mn$, $g_{\la\mn}$, and $H_\mn$ 
are the relevant coefficients for Lorentz violation.
The coefficients $a_\mu$, $e_\mu$, and $f_\mu$ are not explicitly included
in Eq.~\refeq{lagr} as they can be removed from the theory in the limit that we consider in this work.\cite{ks1998, Altschulf}
We avoid explicitly introducing the coefficient $c_\mn$
because all of its components have been probed already
by other work.\cite{datatables}
The trace of $d_\mn$ is not Lorentz violating and is also 
not considered here.
The fermion field is denoted $\ps$,
$\ga^\mu$ are the Dirac matrices,
and $m$ is the fermion mass.
In the single-fermion limit here in flat spacetime, there are 44 observable degrees of freedom that are characterized by combinations of the coefficients appearing in the Lagrange density.
Those 44 combinations are conveniently parameterized by the so-called tilde coefficients (see Table P56 of Ref.~\refcite{datatables}).

Some of the most sensitive experiments in the fermion sector\cite{berlin,electron,sensb}
are those that look for spin-precession effects stemming from an interaction described
by the hamiltonian\cite{lane}
\beq
H = \bt_j \si_j,
\eeq
where
the $\bt_j$, three of the 44 tilde coefficients, are the combination of SME coefficients
\beq
\bt_j = b_j -\half\ve_{jkl}H_{kl}-m(d_{jt}-\half\ve_{jkl}g_{klt}),
\label{Eq:btilde}
\eeq
and $\si_j$ are the Pauli spin matrices.
The SME coefficients are typically taken as spacetime constants
in a Sun-centered coordinate system,\cite{datatables}
making them time-dependent in a laboratory frame on Earth
due to the rotation of the Earth at the sidereal frequency $\om$,
the boost of the earth as it orbits the Sun ($\be_\oplus\sim10^{-4}$) at the annual frequency $\Om$,
and the boost of the laboratory as it revolves around the Earth ($\be_L\lesssim10^{-6}$) at the sidereal frequency.
These motions generate periodic signals in experiments at the related frequencies
with amplitudes proportional to the coefficients for Lorentz violation.
Extracting these amplitudes from experimental data is the key measurement technique
in the experiments that we consider here.

Most experiments to date have taken advantage of the rotation of the experiment at the sidereal frequency only,
with some taking advantage of approximate boost effects considered at linear order in the boost velocity (see, e.g., Refs.~\refcite{electron} - \refcite{fgt})
and even fewer considering effects at second order in boost effects.\cite{boostsquare}
Searches for anomalous spin-precession effects are among the most sensitive searches in the fermion sector.  
We have shown that utilizing boost effects beyond linear order
in a reinterpretation of these highly sensitive tests
results in complete coverage of the 44-degree-of-freedom coefficient space.\cite{jof}

In what follows,
we use the the standard SME convention in which lower-case indices denote components of tensors
in the lab frame and upper case indices denote components of tensors in the Sun-centered frame.
To express $\bt_j$ in terms of the constant Sun-centered frame coefficients,
we must use the transformation matrix $A_\mu^{\phantom{j} \Xi}$ that takes us from the Sun-centered frame to the lab frame to transform each of the SME coefficients that make up $\bt_j$.
Hence the following substitutions are used in Eq.~\refeq{Eq:btilde}:
\bea
b_j &=& A_j^{\phantom{j} \Xi} b_\Xi
\label{btran}\\
d_{jt} &=& A_j^{\phantom{j} \Xi} A_t^{\phantom{j} \Si} d_{\Xi \Si}\\
H_{kl} &=& A_k^{\phantom{j} \Xi} A_l^{\phantom{j} \Si} H_{\Xi \Si}\\
g_{klt} &=& A_k^{\phantom{j} \Xi} A_l^{\phantom{j} \Si} A_t^{\phantom{j} \Pi} g_{\Xi \Si \Pi}.
\label{gtran}
\eea
The result can then be rewritten in terms of tilde coefficients in the Sun-centered frame
and Fourier decomposed to show the frequencies at which time dependence of the experimental signals arises.  The amplitudes of the Fourier components
are then proportional to the coefficients associated with each of those signals.
The full transformation matrix is composed as $\mathbf{A}= \mathbf{R \La_L \La_\oplus}$,
where 
\beq
\mathbf{R} = 
\begin{pmatrix}
1 &0 &0 &0\\
0 &\cos \ch\cos\om t &\cos \ch\sin\om t  &-\sin\ch\\
0 &-\sin\om t & \cos\om t & 0\\
0 & \sin \ch\cos\om t & \sin \ch\sin\om t &\cos\ch
\end{pmatrix}
\eeq
is the rotation that takes us to a lab-fix coordinate system with the $z$-axis pointing vertically upward in the lab.
The angle $\ch$ is the colatitude of the experiment,
which we take as zero for simplicity in the examples presented in this work.
The remainder of the transformation involves boost matrices $\mathbf{\La_\oplus}$ and $\mathbf{\La_L}$ of the form
\beq
\mathbf{\La} = 
\begin{pmatrix}
\ga &\ga\be_1 &\ga \be_2 &\ga\be_3\\
\ga\be_1 &1+ \fr{(\ga-1)\be_1^2}{\be^2} & \fr{(\ga-1)\be_1\be_2}{\be^2}  &\fr{(\ga-1)\be_1\be_3}{\be^2}\\
\ga\be_2 &\fr{(\ga-1)\be_1\be_2}{\be^2} & 1+\fr{(\ga-1)\be_2^2}{\be^2}  &\fr{(\ga-1)\be_2\be_3}{\be^2}\\
\ga\be_3 & \fr{(\ga-1)\be_1\be_3}{\be^2} & \fr{(\ga-1)\be_2\be_3}{\be^2}  &1+\fr{(\ga-1)\be_3^2}{\be^2}.
\end{pmatrix}
\eeq
Here $\be_1,\be_2,\be_3$ are the components of the respective boost velocities $\vec \be_\oplus$ and $\vec \be_L$.
In the examples to follow,
we work with a circular orbit approximation for Earth's orbit around the sun,
where  
\beq
\vec \be_\oplus = \be_\oplus (\sin\Om t, -\cos \et \cos \Om t, -\sin \et \cos \Om t),
\eeq
and $\et$ is the inclination of Earth's orbit in the Sun-centered frame.
We also set aside consideration of $\be_L$ effects.

\section{The sidereal frequency and new discovery potential}

It is perhaps surprising that consideration of effects suppressed by the annual boost of the Earth's revolution around the Sun leads to additional signals at the sidereal frequency.  However, this feature indeed occurs and results in new limits on Lorentz violation that can be read from existing published results.
It may be similarly surprising that considering higher boost suppressions leads to 
added discovery potential in ``old" experiments, but this also happens.
In this section we illustrate how these effect arises using a simple limit of the full SME.
We should emphasize that this material is a useful illustration of how more complicated results arise in the full SME and is not intended for experimental analysis, which should instead be pursued in the context of the full study.\cite{jof}

Suppose that $  d_{TX} $ and $  d_{XT} $
are the only nonzero Sun-centered frame coefficients.
In this limit, the only nonzero tilde coefficients\footnote{Note that in this limit $\bt_X^* = -\bt_X$ does not characterize an independent degree freedom.} in the Sun-centered frame are $  \bt_X   = -m d_{XT} $ and $  \dt_X  =  m (  d_{TX}  + \half   d_{XT} )$.
Consider contributions to $\bt_x$ up to order $\be_\oplus^2$ arising from what remains of the calculation described in Eqs.~\refeq{btran}-\refeq{gtran}
\beq
  \tilde b_x   =
- m A_x^{\phantom x X} A_t^{\phantom t T}  d_{XT}    - m A_x^{\phantom x T} A_t^{\phantom t X}  d_{TX}.
\eeq
To make further progress,
we write the rotation and boost parts of this transformation explicitly.
In order to do so, we must introduce the intermediate coordinate system,
one that moves with the Earth but is nonrotating,
denoted with a bar
\bea
\nonumber
\tilde b_x &=& 
- m R_x^{\phantom x \bar x} \Big[(\La_\oplus)_{\bar x}^{\phantom x X}(\La_\oplus)_t^{\phantom t T}  d_{XT} + (\La_\oplus)_{\bar x}^{\phantom x T} (\La_\oplus)_t^{\phantom t X}  d_{TX}\big]\\
& & - m R_x^{\phantom x \bar y} \Big[(\La_\oplus)_{\bar y}^{\phantom x X}(\La_\oplus)_t^{\phantom t T}  d_{XT} + (\La_\oplus)_{\bar y}^{\phantom x T} (\La_\oplus)_t^{\phantom t X}  d_{TX}\Big].
\eea
Inserting the relevant matrix components yields
\bea
\nonumber
\tilde b_x   &=& -m \cos \om t \Big[(1 + \half  \be_\oplus^2)d_{XT} 
+ \be_\oplus^2 \sin^2 \Om t (\half d_{XT} + d_{TX}) \Big] 
\\
& & -m \sin \om t \Big[ 
- \be_\oplus^2 \cos \et \sin \Om t \cos \Om t(\half d_{XT} + d_{TX}) \Big].
\eea
We can express this result in terms of Sun-centered frame tilde coefficients as
\bea
\nonumber
\tilde b_x &=& (1 + \half  \be_\oplus^2) \tilde b_X \cos \om t      \\
& & -   \be_\oplus^2 \tilde d_{X} \cos \om t \sin^2 \Om t   
+ \be_\oplus^2 \tilde d_{X} \cos \et \sin \om t \sin \Om t \cos \Om t,
\eea
which can be decomposed by Fourier component as
\bea
  \bt_x   &=& \Big[ (1 + \half  \be_\oplus^2)   \bt_X   - \half  \be_\oplus^2   \dt_X   \Big] \cos \om t
  \label{Eq:toyFourier}\\
  \nonumber
 && + \quar  \be_\oplus^2   \dt_{X} (1 + \cos \et)   \cos (2\Om t - \om t) 
 + \quar  \be_\oplus^2   \dt_{X} (1 - \cos \et)  \cos (2\Om t + \om t).
 \eea
 Though our work here is in a comparatively simple limit of the full SME,
 the steps carried out above illustrate the same steps used in the full analysis.\cite{jof}

Because the SME is a test framework and not a model,
one must choose how many coefficients to consider at one time in a fit to data.
This is essential given that the full SME is an infinite series.
One common choice that provides a sense of the basic level of depth to which a coefficient has been probed
is a Maximum-Reach Analysis\cite{fgt} in which each Lorentz-violating degree of freedom is fit to data separately under the assumption that all of the other coefficients are zero.
Applying this approach in the context of Eq.~\refeq{Eq:toyFourier}
plays out as follows in the neutron sector.
Prior analysis using the $\be_\oplus=0$ limit
placed the constraint $\bt_X < 10^{-33}$~GeV by searching for the $\cos \om t$ Fourier component,\cite{berlin}
while being insensitive to $\dt_{X}$.
Here we see that in a Maximum-Reach Analysis, setting $\dt_{X}=0$ and considering $\bt_x$ measurements generates
the same result as existing studies from the $\cos \om t$ Fourier component,
while the reverse, setting $\bt_X=0$ and constraining $\dt_{X}$ using the $\cos \om t$ Fourier component,
results in a limit of $\dt_{X}<10^{-25}$~GeV.
Hence we demonstrate that by considering additional powers of the boost velocity,
a Maximum-Reach Analysis results in additional bounds on new coefficients using existing published results established by searching for sidereal variations.

Because the original experiments were so highly sensitive,
the new bound we achieve after suffering a suppression of eight orders of magnitude
is still arguably beyond the level one might expect for Planck-suppressed physics.
For a coefficient with units of GeV as we have here,
we might, as one possibility,
generate a crude dimensional argument for the scale of Planck suppressed physics as follows
\beq
\fr{m_n^2}{m_{\rm Planck}} \sim 10^{-19}\, {\rm GeV},
\eeq
where $m_n\sim 1$ GeV is the neutron mass and $m_{\rm Planck} \sim 10^{19}$ GeV is the Planck mass.

Another common approach in the literature is to consider fitting some larger subset of SME coefficients
to data together.  This is known as a Coefficient-Separation Approach.\cite{fgt}
In the context of the current example,
suppose we consider $\bt_X$ and $\dt_X$ together.
This approach will require two independent measurements of different linear combinations of $\bt_X$ and $\dt_X$
to obtain independent measurements of both coefficients.
In principle this can be done by searching for the $\cos \om t$ Fourier component
along with at least one of the $\cos (2\Om t \pm \om t)$ Fourier components.
This illustrates that the methods we describe here open additional discovery potential for
new analysis and/or experiments.
Here one could, for example, imagine the possibility that $(1 + \half  \be_\oplus^2)\bt_X = \half  \be_\oplus^2   \dt_X$.  Thus Lorentz violation would be absent in searches at the sidereal frequency but observable at $2\Om t \pm \om t$.
In the context of this example,
some of these features may be difficult to observe
because the frequencies are close together, requiring a large amount of date to separately observe the amplitudes.
The point is to illustrate how some of the features seen in a more comprehensive approach\cite{jof}
arise in the context of a tractable example.

\section{The $g_{\la \mn}$ coefficients}

There are many components of $g_{\Xi \Si \Pi}$
that have not yet been experimentally explored\cite{datatables}
that we succeed in accessing with higher-precision boost transformations.\cite{jof}
It may be surprising that higher-precision boost transformations
generate such access.
This is a second feature of our work that is interesting
to explore in the context of a simplified example.
Let us take the example of $\gt_{TX} = m (g_{YTZ} + g_{ZTY})$, which is one such unexplored coefficient.
A simple limit of the full SME in which this coefficient is the only nonzero tilde coefficient
in the Sun-centered frame is to consider $g_{YTZ}= g_{ZTY} = - g_{TYZ} = - g_{TZY}$ as the only nonzero SME coefficients.

In this context, the nonzero contributions to $\bt_x$ can be written
\beq
\bt_x = m g_{yzt},
\eeq
or in terms of Sun-centered frame coefficients
\bea
\nonumber
\bt_x &=& m A_y^{\phantom{x}\Xi} A_z^{\phantom{x}\Si} A_t^{\phantom{x}\Pi} g_{\Xi \Si \Pi}\\
&=& 
\nonumber
m A_y^{\phantom{x}Y} A_z^{\phantom{x}T} A_t^{\phantom{x}Z}  g_{YTZ}
+ m A_y^{\phantom{x}T} A_z^{\phantom{x}Z} A_t^{\phantom{x}Y} g_{TZY}\\
&& + m A_y^{\phantom{x}Z} A_z^{\phantom{x}T} A_t^{\phantom{x}Y}  g_{ZTY}
+ m A_y^{\phantom{x}T} A_z^{\phantom{x}Y} A_t^{\phantom{x}Z} g_{TYZ}.
\label{gstuff}
\eea
Let us focus on observables that appear as the $\cos \om t$ Fourier component
of the signal.
The $\cos \om t$ Fourier components in Eq.~\refeq{gstuff} arise from
the transformation matrices of the form $A_y^{\phantom{x}\Xi}$
that appear in each term.
This factor can be expanded as follows
\beq
A_y^{\phantom{x}\Xi} =
R_y^{\phantom{x}\bar y} (\La_\oplus)_{\bar y}^{\phantom{x}\Xi}
+R_y^{\phantom{x}\bar x} (\La_\oplus)_{\bar x}^{\phantom{x}\Xi}.
\label{layxi}
\eeq
We will focus here on the first term in Eq.~\refeq{layxi},
setting aside the second term which produces $\sin \om t$.
The rotation factors in the other $A_\mu^{\phantom{x}\Xi}$ appearing in Eq.~\refeq{gstuff} are one.

Inspection of the second line of Eq.~\refeq{gstuff} reveals that all three of the boost matrix components are off diagonal, meaning that these terms generate contributions beyond quadratic order in $\be_\oplus$
while the first line involves one on-diagonal and two off diagonal boost matrix components.
Thus the following quadratic contributions to $\bt_x$ involving $\cos \om t$
arise
\beq
\bt_x \subset m \cos \om t \Big[(\be_\oplus)_Z^2 g_{YTZ} + (\be_\oplus)_Y^2 g_{TZY}\Big].
\eeq
Using the antisymmetry of $g_{\la \mn}$ on the first two indices
and the explicit form of the boost components yields
\beq
\bt_x \subset m \cos \om t \be_\oplus^2 (\sin^2 \et g_{YTZ} - \cos^2 \et g_{ZTY}) \cos^2 \Om t.
\eeq
Using the defining equality of the special limit that we consider here,
limiting attention to the $\cos \om t$ Fourier component, and rewriting in terms of $\gt_{TX}$, we find
\beq
\bt_x \subset - \quar \cos \om t \be_\oplus^2 \cos 2 \et \gt_{TX},
\eeq
revealing that indeed our sample unmeasured $g_{\Xi \Si \Pi}$-containing tilde coefficient,
$\gt_{TX}$, does appear in $\bt_j$ searches,
and in fact does so at the sidereal frequency at two boost supressions.

\section{Discussion}

In this work we provide some examples
of the kinds of access to coefficients that can be gained 
when considering higher-precision boost transformations.
We show that new combinations of SME coefficients can be accessed at the commonly used sidereal frequency and
that the search for variations at additional frequencies
provides additional discovery potential.
We have also demonstrated that components of the $g_{\la \mn}$ coefficient
that have never been observed arise in a higher-precision boost analysis in a relatively
straightforward way.
These examples are intended to provide an enhanced conceptual understanding
of the results that appear in our more comprehensive work.\cite{jof}

There are other interesting features that emerge in a higher-precision boost analysis beyond those highlighted in the examples provided here.
One notable case is the accessibility of the $\bt_J^*$ coefficient
that is typically thought of as ripe for exploration with antimatter,
as has been done for the proton.\cite{protonstar}
Suppressed access to this coefficient by our methods complements existing approaches to exploring this coefficient with ordinary matter.\cite{nr}
A special limit of the SME can be constructed in which $\bt_J^*$ is the only
nonzero Lorentz-violating degree of freedom.
This forms an interesting example for potential future work,
though its complexity may limit its value as an illuminating example.

In this work we focus our examples on the simplified limit of a circular orbit
for the earth.  
In our comprehensive treatment, 
we consider a more complicated model of Earths motion including 
effects such as the ellipticity of the Earth's orbit
in the context of the full coefficient space for protons, neutrons, and electrons.
We also consider implications for the muon sector in other work.\cite{muon1}
The consideration of higher-precision boost transformations
is a surprisingly powerful and versatile tool in the ongoing search for Lorentz violation.

\end{document}